# eMZed 3: flexible and interactive development of scalable LC-MS/MS data analysis workflows in Python

**Uwe Schmitt[1*], Jethro L. Hemmann[2*], Nicola Zamboni[3], Julia A. Vorholt[4], Patrick Kiefer[4*]**

*****Equal contribution

[1]Scientific IT Services, ETH Zurich, Zurich, Switzerland

[2]Junior Research Group Metabolomics-guided Natural Product Discovery, Leibniz Institute for Natural Product Research and Infection Biology (Leibniz-HKI), Jena, Germany

[3]Institute of Molecular Systems Biology, ETH Zurich, Zurich, Switzerland

[4]Institute of Microbiology, Department of Biology, ETH Zurich, Zurich, Switzerland

E-mail: pkiefer@ethz.ch

## Abstract

**Summary:** Liquid chromatography–mass spectrometry (LC-MS/MS) data analysis requires adaptable software solutions to meet diverse analytical needs. We present eMZed 3, a modern Python framework for flexible and interactive analysis of LC-MS/MS data. eMZed 3 enables users to develop scalable workflows tailored to their specific requirements while leveraging Python's extensive ecosystem of libraries. Building on its predecessor, eMZed 3 is now Python 3-based and includes substantial enhancements, including support for chromatogram-based LC-MS data, a new SQLite-based backend supporting optional out-of-memory processing, and rich interactive visualization tools. Compared to the previous version, eMZed 3 is now split into three packages: *emzed* (core functionalities), *emzed-gui* (interactive data visualization), and *emzed-spyder* (an integrated development environment). This modular architecture allows straightforward integration of the *emzed* core library into headless Python environments, including computational notebooks (such as Jupyter) or high-performance computing clusters. eMZed 3 incorporates well-established libraries such as OpenMS, and is suited for both targeted and untargeted metabolomics. Overall, eMZed 3 supports the efficient development of scalable and reproducible LC-MS data analysis and is accessible to both novice and advanced programmers.

**Availability and Implementation:** eMZed 3 and its documentation are freely available at https://emzed.ethz.ch, the source code is hosted at https://gitlab.com/groups/emzed3.
An online-executable example workflow is available on Binder at:
https://mybinder.org/v2/gl/emzed3%2Femzed-example-workflow/HEAD?labpath=example.ipynb.

**Contact:** pkiefer@ethz.ch

## Introduction

Liquid chromatography–mass spectrometry (LC-MS) is a core analytical technique in many fields of biology, environmental sciences, and chemistry. Advances in LC-MS instrumentation have enabled the generation of increasingly large and complex data sets, leading to computational challenges. Consequently, reliable and reproducible analysis of LC-MS data has become a critical step in the analytical workflow and is supported by a variety of commercial and open-source software tools. These tools often belong to one of two extremes: user-friendly software with limited flexibility, and libraries that require advanced programming skills but offer greater versatility. For example, MZmine 3 (Schmid *et al.* 2023) is a widely-used GUI-based application allowing the simple setup of data analysis workflows consisting of pre-defined processing steps. However, it offers limited flexibility and lacks easy programmatic customization. In contrast, the OpenMS project (Pfeuffer *et al.* 2024) provides a robust and performant C++ library with Python bindings that allows high flexibility in workflow programming, but lacks advanced interactive visualization capabilities. To bridge this gap and support the interactive prototyping and development of tailored workflows, we had introduced eMZed as a novel open-source Python framework (Kiefer, Schmitt, and Vorholt 2013). Since then, eMZed has been used within a number of research projects (Hartl *et al.* 2020, Martano *et al.* 2020, Nguyen *et al.* 2020, Reiter S *et al.* 2020, Hemmann *et al.* 2021, Maier *et al.* 2021, Butin *et al.* 2022, Keller *et al.* 2022, Ryback, Bortfeld-Miller, and Vorholt 2022, Reiter MA *et al.* 2024, Mori *et al.* 2025) and continued to be further developed.

Here, we present a completely revised and modernized version 3 of eMZed. eMZed 3 has been rewritten for compatibility with Python 3 (Van Rossum and Drake 2009) and features numerous new functions and performance enhancements, including support for chromatogram-based LC-MS/MS data (e.g. multiple reaction monitoring data), out-of-memory processing, and improved interactive data inspection and visualization. The framework retains its core strength of offering a comprehensive set of elementary functions and data types for reading, processing, and analyzing LC-MS data. With an intuitive Python API, eMZed 3 allows users to create custom workflows that seamlessly integrate with popular Python libraries, such as *pandas* (McKinney 2010), *NumPy* (Harris *et al.* 2020), or *scikit-learn* (Pedregosa *et al.* 2011). It also enables interactive visualization of chromatograms, spectra, and peakmaps at any stage of a workflow. The optional Spyder-based integrated development environment (IDE)[1] bundles the eMZed modules, helping programming beginners to quickly develop their own LC-MS data analysis workflows. eMZed supports all aspects of reproducible and tailored data analysis for both targeted and untargeted metabolomics, as well as other applications.

## Results

### Functionalities

eMZed was designed as an easy-to-use coding framework that allows users to develop highly customizable workflows for LC-MS data analysis in Python (**Fig. 1**). To that end, eMZed builds upon existing algorithms from e.g. PyOpenMS (Röst *et al.* 2014) while incorporating additional features, including a versatile table data structure to store and access LC-MS peak lists and GUI-tools for the interactive inspection and re-integration of peaks. While eMZed includes many features found in widely used tools such as MZmine (Schmid *et al.* 2023) and XCMS (Louail *et al.* 2025) strength lies in enabling users to program custom workflows beyond the scope of existing tools and to address specific problems requiring tailored solutions (see also example workflow below).

The new version 3 of eMZed builds on the architecture and core features of the previous version of eMZed (Kiefer, Schmitt, and Vorholt 2013), but is now compatible with Python 3 and fully separated into three individual packages (all available at PyPI): *emzed*, *emzed-gui*, and

[1] https://www.spyder-ide.org/

*emzed-spyder*. The package *emzed* provides core functionalities of eMZed, while the optional *emzed-gui* allows interactive data inspection and visualization (see **Fig. 2A** for example code). To facilitate the interactive development of workflows with eMZed, *emzed-spyder* bundles a customized version of the Spyder IDE with the eMZed packages. For Windows, an installer is available at the eMZed website that allows an easy installation of eMZed 3 (*emzed-spyder* with all other eMZed packages).

The Python package *emzed* offers the core functionalities of eMZed 3, and does not depend on any GUI components. Thus, *emzed* can be used with any type of Python installation, including high-performance computing clusters or Jupyter notebooks (Kluyver *et al.* 2016). Among many features, *emzed* provides an easy-to-use API for:

- Direct access to chromatograms, peaks, and spectra ($MS^n$) and their underlying data
- Targeted peak extraction
- Untargeted feature finding, isotope, and adduct grouping
- Alignment of peaks (m/z and retention time) and feature lists
- Powerful table-based operations on peak lists
- Handling of chemical data such as molecular formulas
- Accessing R functionalities using integrated bridge to R

Several integration algorithms are provided by *emzed* and allow fast, parallel integration of chromatographic peaks. For untargeted analysis, eMZed integrates the feature finding algorithms FeatureFinderMetabo from OpenMS (Röst *et al.* 2016) and centWave (Tautenhahn, Böttcher, and Neumann 2008), which is accessible from *emzed* via a builtin bridge to R (R Core Team 2021). Additionally, the ADAP feature finder from MZmine 2 (Myers *et al.* 2017) is available as an eMZed extension. eMZed 3 now also supports chromatogram-based LC-MS data, e.g. MRM data from triple quadrupole instruments. The powerful *Table* data structure plays a key role in eMZed, allowing operations such as filtering, grouping, joining, and aggregating and it now uses an SQLite database (Gaffney *et al.* 2022) as backend for improved performance and optional out-of-memory operations for processing large data sets.

For interactive visual inspection of *emzed* data structures, the *emzed-gui* package provides several tools that have been improved in eMZed 3. The *Peakmap Explorer* allows quick exploration of entire LC-MS/MS peakmaps using a 2D heatmap for visualization and can display chromatograms as well as mass spectra (**Fig. 22B**). The *Table Explorer* is a central element of eMZed and assists in inspecting and modifying data stored in *emzed* tables (**Fig. 2C**). A prominent use case for a *Table* is the management of peak lists, and in this case the *Table Explorer* provides interactive visualizations and modifications of extracted ion chromatograms and mass spectra (e.g. for manual re-integration of peaks). In eMZed 3, *Table Explorer* now offers improved speed and robustness and more features, e.g. the coloring of individual rows or cells. *emzed-gui* also comes with tools to create simple graphical input forms that can be used in a workflow for user input.

The optional *emzed-spyder* package is a customized IDE to facilitate interactive workflow development and prototyping. *emzed-spyder* is based on Spyder 5 and integrates all other eMZed packages. It supports direct inspection of *PeakMap* and *Table* data structures in the variable explorer and includes functionalities for the development of eMZed-based Python projects (also called *eMZed extensions*) based on separate virtual environments created for that purpose.

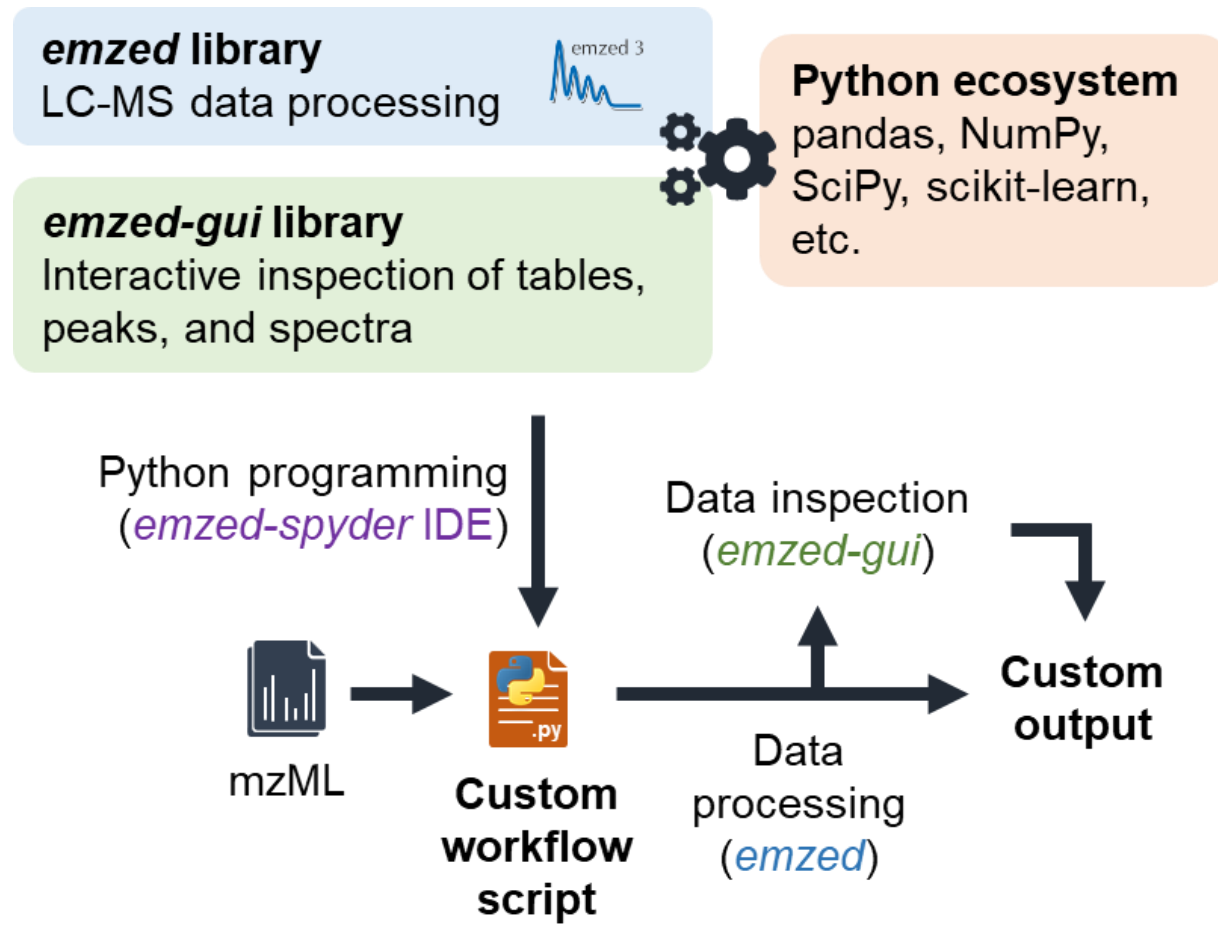

**Fig. 1**: **The core concept of eMZed.** eMZed is designed as a modular framework (*emzed*, *emzed-gui*, *emzed-spyder*) to enable the Python-based development of highly customizable processing workflows for LC-MS/MS data. It incorporates key functions to access and process peaks, spectra, and chromatograms and connects with commonly used libraries from the Python ecosystem.

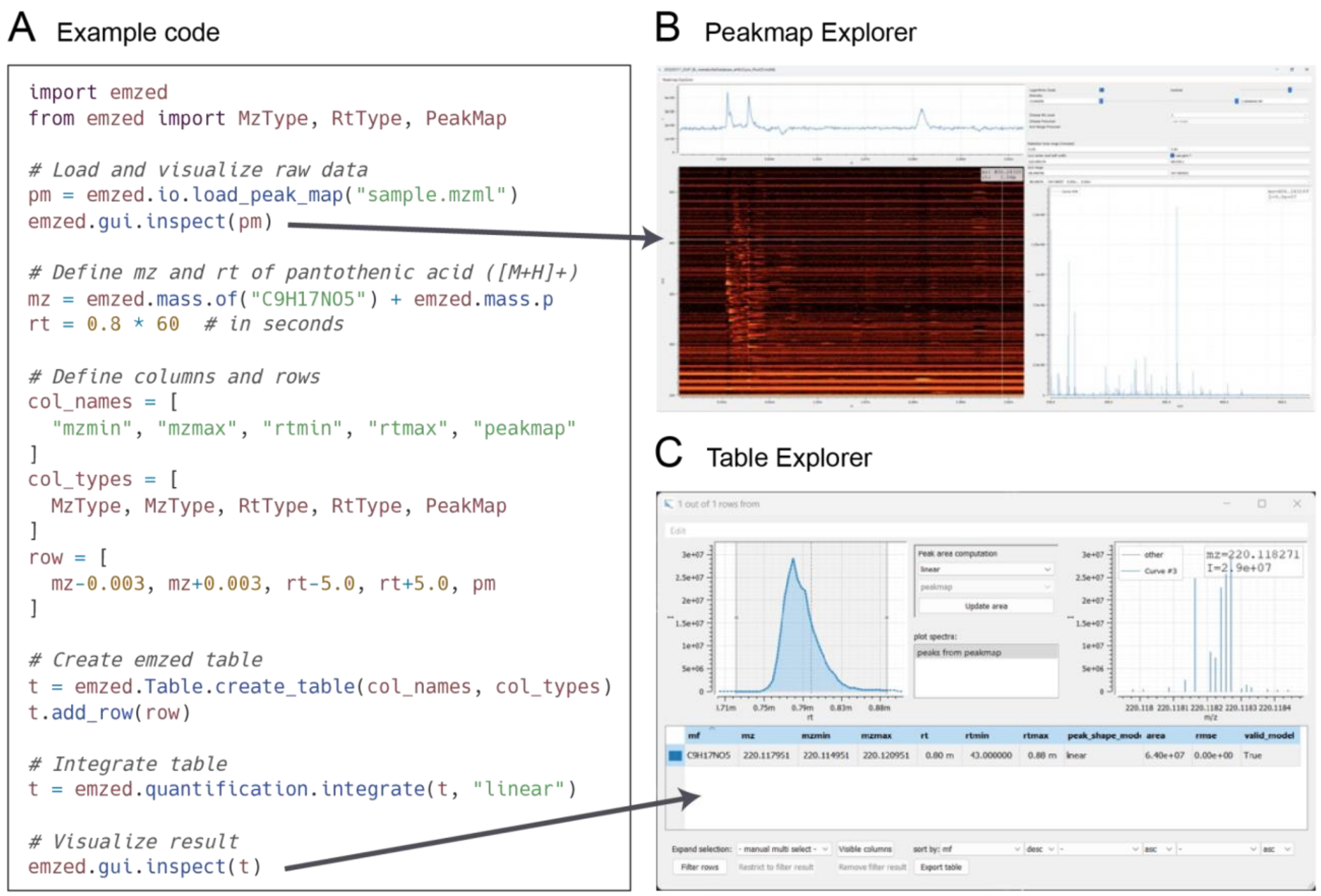

**Fig. 2**: **Visualization of LC-MS data using eMZed 3.** *A*, Example code demonstrating a typical workflow: loading of LC-MS data, inspection of the peakmap, extraction and integration of LC-MS peaks corresponding to specific metabolites, and peak visualization. *B*, Screenshot of *Peakmap Explorer*, which allows the interactive inspection of LC-MS/MS peakmaps. *C*, Screenshot of *Table Explorer*, used to interactively inspect and (re-)integrate LC-MS peaks which are defined in an eMZed *Table* by their *m/z* and retention time values.

## Implementation details

For the development of eMZed 3, the previous Python 2-based code (Kiefer, Schmitt, and Vorholt 2013) was revised and modernized to be compatible with Python 3 (tested with versions 3.11−3.14). While most of the architecture from the previous version of eMZed was kept, the new implementation includes several notable technical improvements. To store and process information within the eMZed data structures *Table* and *PeakMap*, an SQLite database is now used internally, which results in a major performance enhancement and significantly reduced memory footprint. The SQLite database is per default kept in memory but optionally allows out-of-memory operations on disk, thus circumventing working memory limitations that might occur for large data. To allow access to functionalities from OpenMS (e.g. I/O operations, feature finding, spectral processing, retention time alignment), PyOpenMS (version 3.3.0) is used, which is run in its own process to prevent version conflicts between shared libraries, such as the Qt framework. The ADAP peak picker from MZmine 2 (Myers *et al.* 2017) is offered as a separate extension module. The GUI tools of eMZed are now migrated to PyQt5. The customized IDE *emzed-spyder* is based on Spyder (version 5.5.6) and enhances Spyder to support eMZed-specific features and offers functions for the development of eMZed-based projects.

eMZed 3 runs on all modern operating systems (tested on Windows, Ubuntu, macOS), and the core library *emzed* is compatible with (cloud-based) notebook environments (e.g. Jupyter, Google Colab).

### Example workflow

To assist new users, we provide a comprehensive tutorial for beginners on the eMZed website (https://emzed.ethz.ch). In addition, we created a more advanced example workflow to demonstrate the capabilities of eMZed 3. The example workflow can be executed as a cloud-based notebook on Binder[2] and highlights how eMZed can be used to correct for retention time (RT) shifts using a problem-specific approach. In this example, we consider the analysis of amino acids in a previously published data set comprising 200 samples (Maier *et al.* 2021). Raw data visualization reveals RT instability within a smaller sample subset. However, simply enlarging the RT windows used for peak integration is not a suitable approach to cope with the observed RT shifts, since the RT difference of the two mass isomers leucine and isoleucine is smaller than the observed maximal inter-sample RT shift. The example workflow thus implements a simple, case-specific solution that fixes the issue by using RT shifts of other amino acids as estimates to shift the RT windows for leucine and isoleucine. In the example notebook, we further evaluate whether the strategy was successful and visualize treatment-specific differences of the amino acid profiles.

## Discussion

eMZed seeks to fill a gap between other commonly used open-source tools for LC-MS data analysis, such as MZmine (Heuckeroth *et al.* 2024), OpenMS (Pfeuffer *et al.* 2024), MetaboAnalyst (Pang *et al.* 2024), or XCMS (Louail *et al.* 2025) by providing a powerful set of low-level functions to access, process, and interactively visualize LC-MS data that seamlessly integrates into the Python ecosystem of libraries for data analysis, machine learning, and plotting. By combining these functionalities, users can quickly develop tailored LC-MS data analysis workflows, while staying close to the raw data (chromatograms and spectra) with the help of the interactive GUI tools. We envision that eMZed 3 will be of particular benefit to users who require customized, yet robust and scalable data analysis pipelines, e.g. for analytical core facilities. We aim to continuously develop eMZed and provide additional eMZed-based tools and workflows in the future. As an open-source software, we particularly welcome community contributions (in the Gitlab repository).

[2] https://mybinder.org/v2/gl/emzed3%2Femzed-example-workflow/HEAD?labpath=example.ipynb

## Acknowledgements

We acknowledge Andrea Zamuner for testing eMZed 3 and providing input for improvements.

## Funding information

This work was supported by ETH Zurich, Department of Biology, within the frame of an IT-strategy initiative, the Personalized Health and Related Technologies (PHRT) initiative of the ETH Domain, as well as the NCCR Microbiomes by the Swiss National Science Foundation (51NF40_180575 and 51NF40_225148). J.L.H. was supported by the Free State of Thuringia and the European Social Fund Plus (2024 FGR 0017).

## Conflict of interests

The authors declare to have no competing interests or other interests that might be perceived to influence the interpretation of the article.